\documentclass{pasj00}
\draft
\SetRunningHead{H. Nakajima et~al.}{X-Ray Observation on the Monoceros R2 Star-Forming Region with the Chandra ACIS-I Array}
\title{X-Ray Observation on the Monoceros R2 Star-Forming Region with the Chandra ACIS-I Array}
\author{Hiroshi \textsc{Nakajima}, Kensuke \textsc{Imanishi}, Shin-ichiro \textsc{Takagi}, Katsuji \textsc{Koyama}}
\affil{%
   Department of Physics, Graduate School of Science, Kyoto University, Sakyo-ku, Kyoto 606-8502}
\email{nakajima@cr.scphys.kyoto-u.ac.jp, kensuke@cr.scphys.kyoto-u.ac.jp, takagi9@cr.scphys.kyoto-u.ac.jp, \\
koyama@cr.scphys.kyoto-u.ac.jp}
\and
\author{Masahiro \textsc{Tsujimoto}}
\affil{Department of Astronomy \& Astrophysics, Pennsylvania State University, \\
525 Davey Lab., University Park, PA 16802, USA}
\email{tsujimot@astro.psu.edu}
\Received{2002/11/29}
\Accepted{2003/4/28}
\Published{2003//}
\begin{document}
\maketitle
%%%%%%%%%%%%%%%%%%%%%%%%%%%%%%%%%%%%%
% Abstract
%
\begin{abstract}

We report on the results of the Chandra ACIS-I observation
on the central region of the Monoceros R2 cloud (Mon R2),
a high-mass star-forming region (SFR) at a distance of 830~pc.
With a deep exposure of $\sim$100~ks, we detected 368 X-ray sources,
$\sim$80\% of which were identified with the near-infrared (NIR) counterparts.
We systematically analyzed the spectra and time variability
of most of the X-ray emitting sources and provided a comprehensive X-ray
source catalog for the first time.
Using the $J$-, $H$-, and $K$-band magnitudes of the NIR counterparts,
we estimated the evolutionary phase (classical T Tauri stars and
weak-lined T Tauri stars) and the mass of the X-ray emitting sources, and
analyzed the X-ray properties as a function of the age and mass.
We found a marginal hint that
classical T Tauri stars have a slightly higher temperature (2.4~keV)
than that of weak-lined T Tauri stars (2.0~keV).
A significant fraction of the high- and intermediate-mass sources
have a time variability and high plasma temperatures (2.7~keV)
similar to those of low-mass sources (2.0~keV).
We performed the same analysis for other SFRs, the
Orion Nebula Cluster and Orion Molecular Cloud-2/3, and obtained
similar results to Mon R2.
This supports the earlier results of this observation obtained by
Kohno et~al. (2002, ApJ, 567, 423) and Preibisch et al. (2002, A\&A, 392, 945)
that high- and intermediate-mass young stellar objects emit X-rays
via magnetic activity.
We also found a significant difference in the spatial distribution
between X-ray and NIR sources.
\end{abstract}

%%%%%%%%%%%%%%%%%%%%%%%%%%%%%%%%%%%%%
% Introduction
%
\section{Introduction}

Low-mass young stellar objects (YSOs) are classified
along their evolutionary phases into classes 0, I, II, and III
based on their infrared (IR) to sub-millimeter spectral energy distributions
(Lada 1991; Andr$\acute{\rm e}$ et~al. 1993;
Andr$\acute{\rm e}$, Montmerle 1994).
Class 0 and I sources correspond to the youngest and evolved protostars, while
class II and III sources roughly correspond to classical T Tauri stars (CTTSs)
and weak-lined T Tauri stars (WTTSs), respectively.
Low-mass TTSs (=CTTSs and WTTSs) were first discovered to emit significant
X-rays with the Einstein observatory
(Feigelson, DeCampli 1981; Feigelson, Kriss 1981; Montmerle et~al. 1983).
The following observations with ASCA and ROSAT found X-rays from
class I sources (Koyama et~al. 1994; Casanova et~al. 1995; Koyama et~al. 1996;
Kamata et~al. 1997; Grosso et~al. 1997, 2000; Tsuboi 1999; Tsuboi et~al. 2000;
Grosso 2001).
In addition, Tsuboi et~al. (2001) reported on the hard X-ray emission from
class 0 candidates using the Chandra X-ray Observatory (CXO).
Thus, it is now widely known that
low-mass YSOs commonly emit X-rays during most of their evolutionary phases.
Although it is not clear for class 0 sources, X-ray emission from most YSOs
is explained by a thermal plasma with temperatures of 1 to a few~keV
and rapid time variances, which are similar to those of the solar X-rays.
Accordingly, the X-ray origin is generally thought to be solar-type
magnetic activity: amplification of the magnetic field followed by
magnetic reconnection associated
with the stellar convection layer and differential rotation
(Feigelson, DeCampli 1981; Montmerle et~al. 1983).
However, whether or not the X-ray emission differs systematically
along the evolutionary phases and stellar masses is not yet known.

On the other hand, the X-ray emission of high-mass main-sequence (MS) stars
is described by a thin thermal plasma model with relatively soft temperatures
of less than 1~keV, and shows moderate variability.
Lucy and White (1980) proposed that these features are due to the
plasma heating attributable to the collision of the accelerated stellar wind
with circumstellar matter.
However, our understanding of X-ray emission from high- and intermediate-mass
YSOs is largely behind the low-mass sources and high-mass MS stars,
because the former evolve very quickly and the population is small; hence,
they constitute fewer or more distant samples than those of low-mass YSOs.
Since YSOs are generally located
in dense cloud cores, not only optical/near-infrared
(NIR) light but also soft X-rays suffer large extinction.
Accordingly, hard X-ray imaging instruments are essential to 
resolve embedded YSOs in the cores.
The CXO (Weisskopf et~al. 2002), with a high spatial resolution, large
field of view (FOV), wide energy range and moderate energy resolution,
provides the best instruments for observing the star-forming
regions (SFRs).

The Monoceros R2 cloud (hereafter Mon R2) is one of the closest high-mass SFRs,
which is located at a distance of 830~pc (Racine 1968).
The cloud age is $\sim$1~Myr (Carpenter et~al. 1997), providing numerous
YSOs over a wide range of ages from protostars to WTTSs.
The core mass is $\sim$1000 $\MO$ (Tafalla et~al. 1997), which is two orders
of magnitude larger than those of the cores of low- to intermediate-mass SFRs.
Thus, Mon R2 is a good candidate for a systematic study of
the X-ray emission along the evolutionary phases and a study of
high-mass star formation.

The first X-ray observation on Mon R2 was made with 
the ROSAT satellite (Gregorio-Hetem et~al. 1998).
However, the spatial resolution was limited to resolve individual YSOs
in the Mon R2 core. Also, a short exposure ($\sim$9500 s) observation
and poor sensitivity in the hard energy band limited the possibility
to detect most of the
embedded YSOs.
Then, the ASCA observation on Mon R2 (Hamaguchi et~al. 2000)
also could not resolve individual YSOs due to the limited spatial resolution
of $\sim 1\arcmin$.
Recently, Kohno et~al. (2002) performed a CXO observation
and resolved X-rays from many YSOs for the first time.
The earlier results have already been reported by Kohno et~al. (2002)
and Preibisch et~al. (2002),
but concentrated only on the central $3\farcm2 \times 3\farcm2$ region.
This paper reports on a follow-up analysis of the CXO data
in the entire cloud core and the NIR cluster of Mon R2.
The purposes of this paper are
(1) to make a comprehensive X-ray catalog,
(2) to discuss the X-ray properties as a function of the evolutionary
phases and the masses of YSOs, and
(3) to discuss the characteristics of X-ray sources by comparing
the spatial distribution of the X-ray and NIR sources.

%%%%%%%%%%%%%%%%%%%%%%%%%%%%%%%%%%%%%
% Observation
%
\section{Observation}

Mon R2 was observed with the Advanced CCD Imaging Spectrometer (ACIS) on board
the CXO on 2000 December 2--4.
The ACIS-I array, comprising four abutted front-illuminated X-ray
CCDs, covered the entire cloud core and the NIR cluster of Mon R2.
The telescope optical axis position on the ACIS-I array was
R.A. = $\timeform{06h07m50s.86}$, Decl. = $\timeform{-06D22'50".0}$
(J2000).
A high spatial resolution of $\sim 0\farcs5$,
a large field of view (FOV) of $17\farcm4 \times 17\farcm4$,
a wide energy range of 0.2--10.0~keV with a moderate energy resolution
($\Delta E\sim$200~eV at $E$ = 6~keV), and a time resolution of 3.2~s
provide us with high-quality and uniform samples of YSOs in the cloud.
We used the level 2 data provided by the
pipeline processing at the Chandra X-ray Center.
Besides, we removed those events
due to charged particles and hot and flickering pixels.
The effective exposure time was then $\sim$100~ks.

%%%%%%%%%%%%%%%%%%%%%%%%%%%%%%%%%%%%%
% Analysis
%
\section{Analysis}

\subsection{Source Detection}

\begin{figure*}[htbp]
\begin{center}
\FigureFile(160mm,160mm){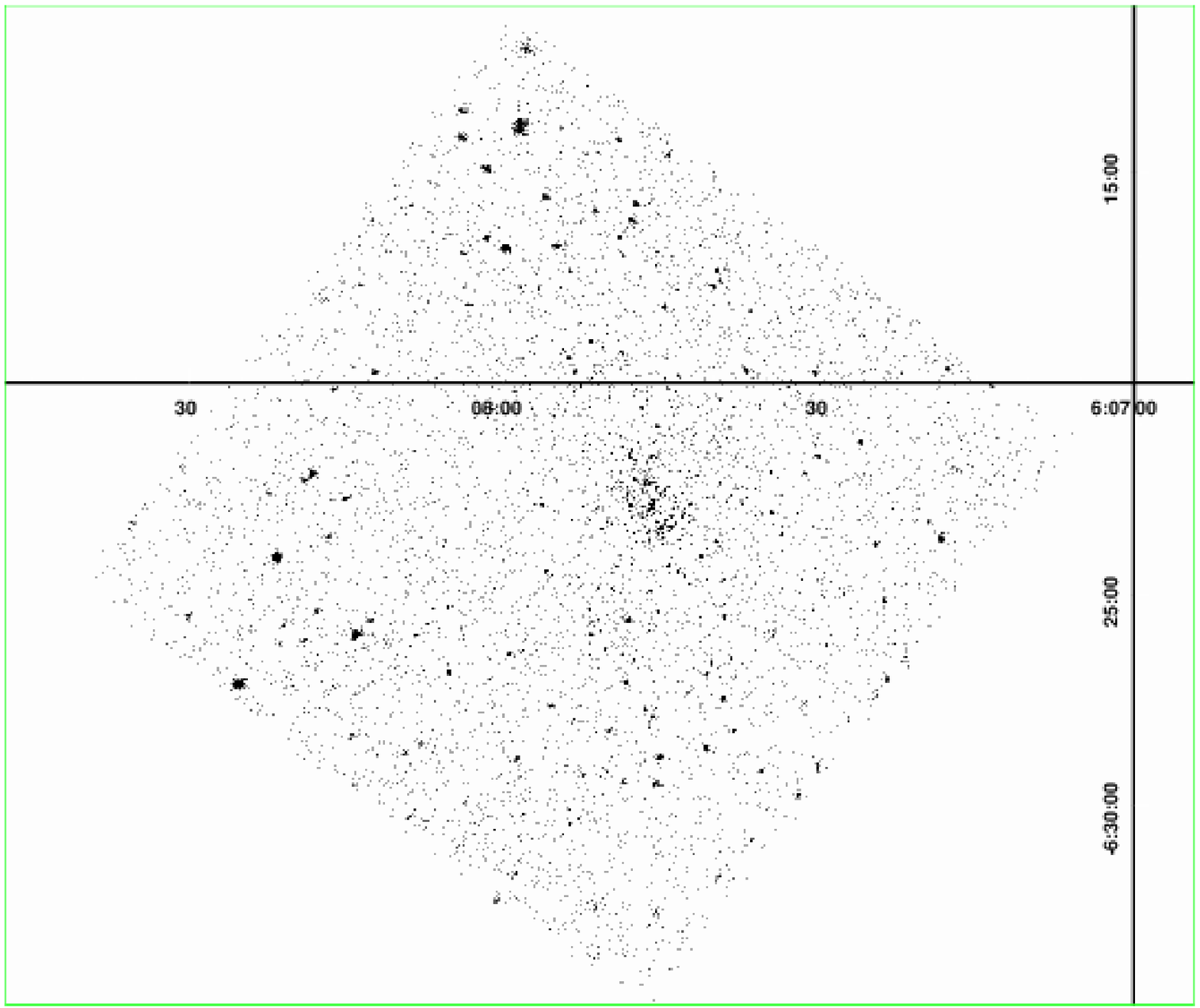}
\end{center}
\caption{ACIS-I image of the entire cloud core
and the NIR cluster of Mon R2 (gray scale) overlaid on the
$^{13}$CO ($J$=1$-$0) intensity contour map
with a beam size of $2\farcm6$ in diameter (Loren 1977).
The square shows the ACIS-I FOV of $17\farcm4 \times 17\farcm4$.
The background photons are extracted from the circular region.
Contours are the antenna temperature at the intervals of 1 K.
The plus indicates the position of the maximum temperature.}
\label{fig:acisI}
\end{figure*}

Figure \ref{fig:acisI} shows an ACIS-I image of the entire cloud core
and the NIR cluster of Mon R2 overlaid on a $^{13}$CO ($J$=1$-$0)
intensity map (Loren 1977).
For source detection, we used the wavdetect program in the CIAO package
version 2.2\footnote{See $\langle$http://asc.harvard.edu/ciao/$\rangle$.}
in the 0.5--10.0~keV band with a threshold significance of
$10^{-6}$ and wavelet scales from 1 to 16 pixels in multiples of $\sqrt{2}$.
In order to study the spectra and timing with reasonable statistics,
we discarded the sources with X-ray counts of less than three-times the
background.
We then detected 368 sources in the ACIS-I FOV.
The X-ray photons of each source were accumulated from a circular region with
a radius of $0\farcs7$--$15\farcs4$,
depending on the point spread function (PSF),
which is a function of the angular distance from the telescope optical axis.
The background photons were extracted from a source-free circle of
$\sim42\arcsec$ in radius.

\subsection{Correlation with the NIR Sources}

To search for the NIR counterparts of the CXO sources,
we used the Point Source Catalog in the 2MASS Second Incremental Data
Release\footnote{See $\langle$http://www.ipac.caltech.edu/2mass/$\rangle$.},
and the results of SQIID and IRCAM3 observations (Carpenter et~al. 1997).
SQIID was mounted at the Cassegrain focus of the 1.3~m telescope in the
Kit Peak National Observatory, which simultaneously takes $J$-, $H$-,
and $K$-band images using 256$\times$256 PtSi arrays (Ellis et~al. 1993).
The pixel scale is $1\farcs36$ providing a FOV of $5\farcm8 \times 5\farcm8$.
The mosaic consists of 3$\times$3 frames, and hence
the SQIID observation obtained a $\sim$15$'\times$15$'$ mosaic image
with the photometric mode centered on Mon R2.
IRCAM3 at UKIRT (United Kingdom Infra-Red Telescope)
utilizes a 256$\times$256 InSb array.
The pixel scale is $0\farcs286$, providing a FOV of $1\farcm2 \times 1\farcm2$.
The observation consisted of 3$\times$3 grid mosaics; hence,
the IRCAM3 observation presented $J$, $H$, $K$, and narrow-band $L$ imaging
photometry of the central $\sim 3\farcm2\times3\farcm2$ region of Mon R2
more deeply than  the SQIID observation.
The 5~$\sigma$ completeness limits in the $K$-band images were 14.5~mag
and 16.6~mag for the SQIID and IRCAM3 observations, respectively
(Carpenter et~al. 1997).
On the other hand, 2MASS has the $K_s$-band limit of 14.3~mag.
Because the position accuracy of 2MASS
(the r.m.s. uncertainty of $\sim0\farcs1$) is better than
those of SQIID ($0\farcs3$) and IRCAM3 ($\sim0\farcs13$),
we used the 2MASS frame as the standard frame.
The distribution of the positional differences between SQIID, IRCAM3, and 2MASS
could be approximated by a two-dimensional Gaussian
in the $\Delta$R.A.--$\Delta$Decl. plane with the center at
$\Delta \rm R.A.\sim 0\farcs65$ and  $\Delta \rm Decl.\sim 0\farcs06$.
We hence fine-tuned the SQIID and IRCAM3 frames so that
the center of the Gaussian would come to the zero position of the 2MASS frame.

We picked up the closest NIR source from each X-ray source and did vice versa.
Using the thus-selected closest pairs of NIR and X-ray sources, we fine-tuned
the ACIS-I frame by shifting $-0\farcs87$ in R.A.
and $-0\farcs75$ in Decl. to match the 2MASS frame.
We then searched for the closest pairs of NIR and X-ray sources again.
Among these pairs, we identified counterpart pairs under the
criterion that the angular distance between the X-ray and the NIR source is
smaller than the root mean square of the position uncertainty
derived by the wavdetect program
and the radius of the PSF.

\begin{figure}[htbp]
\begin{center}
\FigureFile(80mm,60mm){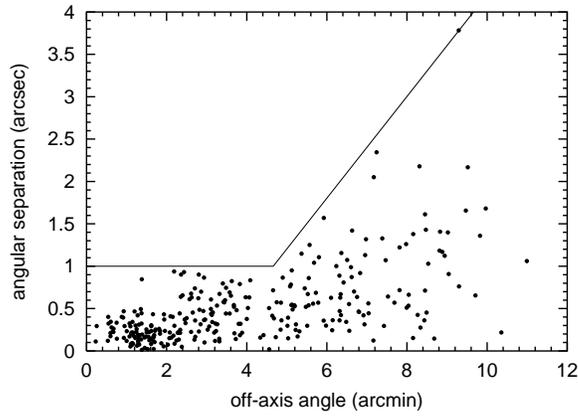}
\end{center}
\caption{Angular separation between X-ray sources
and the NIR counterparts plotted as a function of the off-axis angle.
The identification criterion for the ONC sources is shown by the solid line.}
\label{fig:ang.sep.}
\end{figure}

Finally, 245 X-ray sources were identified with the NIR counterparts in the
SQIID and IRCAM3 databases, and outside of the FOV of SQIID, additional 45
X-ray sources were identified with the 2MASS sources (see figure \ref{fig:ang.sep.}).
Assuming a random distribution of the NIR sources, the chance coincidence
that any of the NIR sources fall in an X-ray source region is estimated to be
$\sim$0.6\%; hence, the false identification is expected to be
less than two sources.
The fine-tuned source positions, background-subtracted photon counts
in the 0.5--10.0~keV band,
and the name of the NIR counterparts are given in table 1.

\subsection{Spectral and Timing Analyses}

\begin{figure}[htbp]
\begin{center}
\FigureFile(80mm,60mm){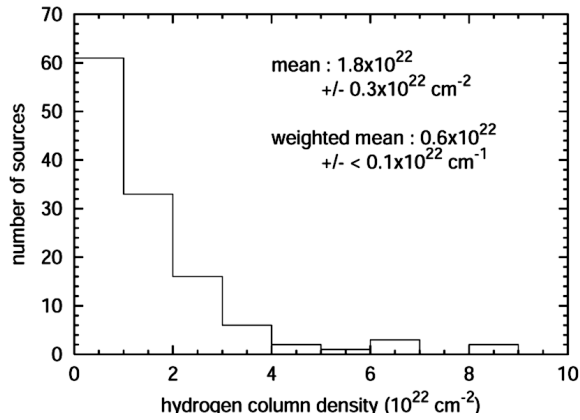}
\end{center}
\caption{Histogram of $N\rm_H$ of the bright sources. The mean value and that weighted with errors of the sources are shown.}
\label{fig:nhdis}
\end{figure}

\begin{figure}[htbp]
\begin{center}
\FigureFile(80mm,60mm){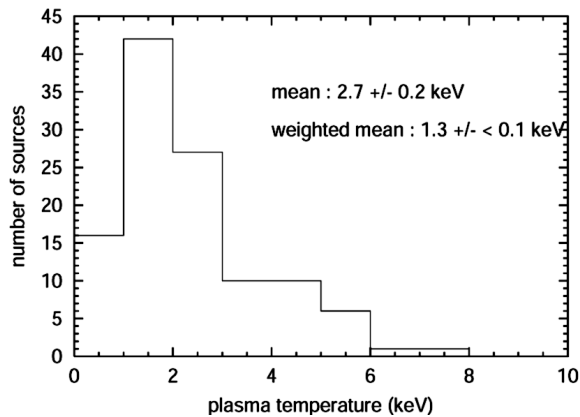}
\end{center}
\caption{Same as figure 3, but for $kT$.}
\label{fig:ktdis}
\end{figure}

\begin{figure}[htbp]
\begin{center}
\FigureFile(80mm,60mm){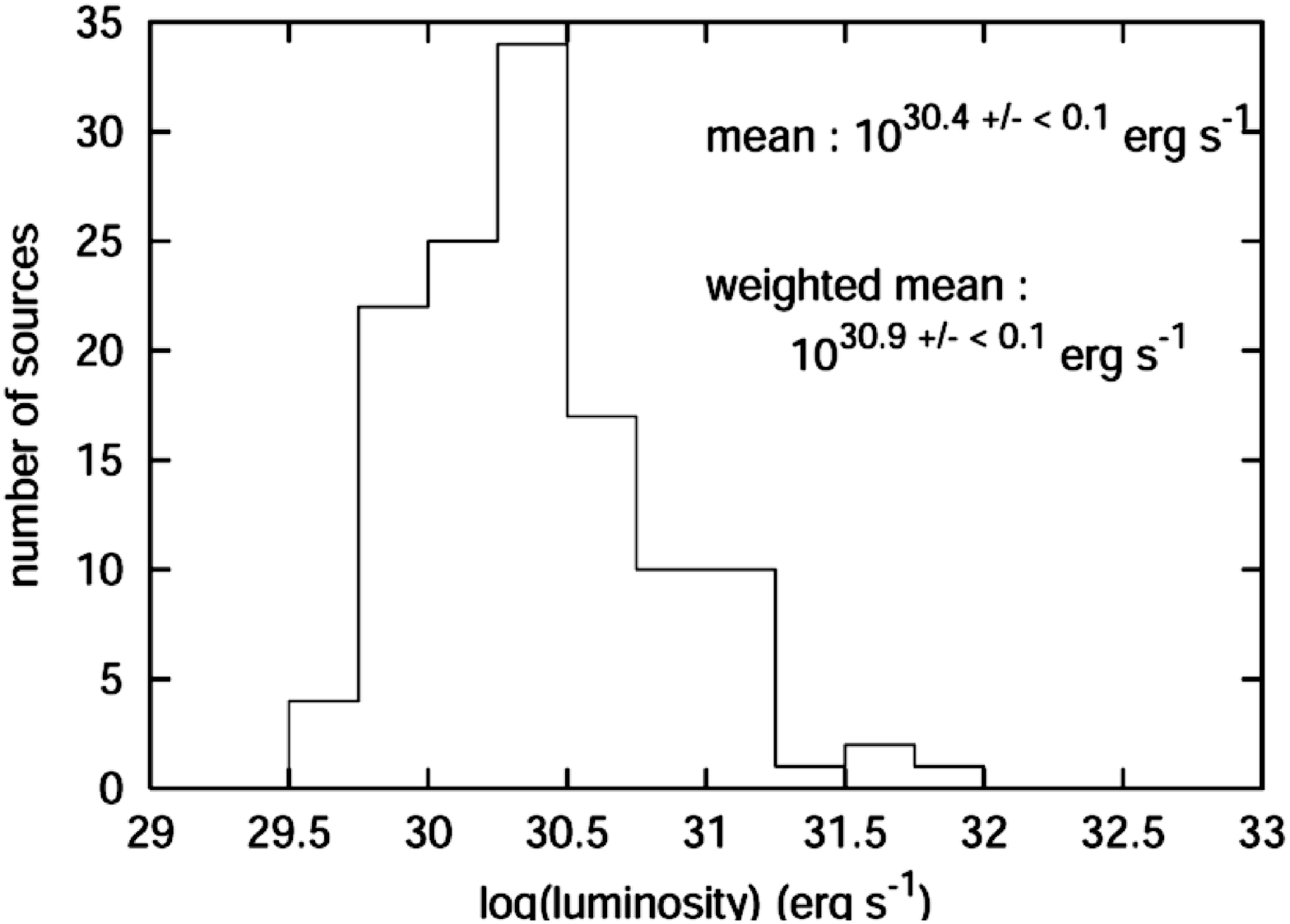}
\end{center}
\caption{Same as figure 3, but for $L\rm_X$.}
\label{fig:ludis}
\end{figure}

For bright 126 sources having more than 50 photons
after background subtraction (hereafter ``bright sources"),
we performed spectral and timing analyses using XSPEC version 10.0 in XANADU.
\footnote{See $\langle$http://heasarc.gsfc.nasa.gov/docs/software/lheasoft/$\rangle$.}
The X-ray spectra were fitted by
a thin-thermal plasma model with three free parameters:
the absorption column density ($N_{\rm H}$), plasma temperature ($kT$),
and normalization.
Since the statistics are limited for most sources, the chemical
abundance was fixed to be 0.3 solar according to previous reports
(Kohno et~al. 2002; Feigelson et~al. 2002; Getman et~al. 2002).
The best-fit parameters and relevant physical values
are listed in table 2 with their 90\% errors in parentheses.
The distributions of $N_{\rm H}$, $kT$,
and $L\rm_X$ of bright sources are shown with
their mean value and that weighed with the errors
(Bevington, Robinson 1992)
in figures \ref{fig:nhdis}, \ref{fig:ktdis}, and \ref{fig:ludis}, respectively.

We then made light curves of the bright sources
in the 0.5--10.0 keV band.
To study the time variability, we applied the Kolmogorov--Smirnov
test (Press et~al. 1988); then, 49 out of 126 sources were found to be
time variables with more than 99\% confidence.
The significance levels of the time variability are listed in table 2.
Most of the variable sources exhibit a flare-like event with
a fast rise and slow decay.

%%%%%%%%%%%%%%%%%%%%%%%%%%%%%%%%%%%%%
% Discussion
%
\section{Discussion}

\subsection{Cloud Members}

We examined whether or not the detected 368 X-ray sources are cloud members
by checking their $N_{\rm H}$ values and $K$-band magnitudes.
The absorption column density of the interstellar medium
to Mon R2 is estimated to be $\sim$2.6$\times10^{21}$~cm$^{-2}$,
assuming the density of the interstellar medium to be 1~cm$^{-3}$.
Since the $N_{\rm H}$ values of most X-ray sources exceed this value
(figure \ref{fig:nhdis}),
they should be located in or behind the cloud,
although the limited photon counts can not give firm constraints.
Most of the background galaxies have $K$ $\gtrsim$ 17~mag
(Tsujimoto et~al. 2002), while the $K$-band limit of NIR sources is 16.6 mag.
We thus suspect that most of the NIR identified X-ray sources are
cloud members.

For all of the ACIS sources, we searched for possible binary systems
by checking whether or not two or more NIR sources were located
within a radius of three-times the PSF.
We used 2MASS and IRCAM3 catalogs,
which have comparable pixel scales of $1\farcs0$ and $0\farcs3$
compared with that of ACIS of $0\farcs5$,
and found that only one source, \#151, has two NIR counterparts separated by
$1\farcs1$ (or $\sim$900 AU).
This separation, however, is one order of magnitude larger than the typical
separation of binaries (Giannuzzi 1989);
hence, these two NIRs may not be a binary system.

\subsection{Comparison with Previous X-ray Observations and Studies}

In order to make a comprehensive discussion about X-ray emitting sources
in Mon R2, we compared our results with those of previous studies.

In a ROSAT/PSPC observation, Gregorio-Hetem et al. (1998) found
23 X-ray point sources in the FOV of ACIS-I.
We searched for ACIS counterparts within a radius of $5\farcs0$ from each PSPC
source (position accuracy $\sim$10$\arcsec$) and found seven counterparts.
We also found five additional possible counterparts relaxing
the criterion radius to the position error of each PSPC source (table 1). 
For the remaining 11 sources, we did not find significant X-rays.
This may have been due to a long-time variability, possibly several years,
as can be seen in the Sun.

There are some sources that are missing in table 1, but are listed
in Kohno et al. (2002).
This discrepancy arises from the difference in the source detection algorithm.
In Kohno et al. (2002), after executing the wavdetect program,
they detected 12 additional sources with a visual
inspection above the 5 $\sigma$ confidence level.
However, in this paper, we did not perform such an inspection, but
excluded 34 sources under the criterion that they should have the source-
significance level of more than three.
Moreover, since we ran the wavdetect only for the total energy band image,
some sources detected only in the soft band (0.5--2.0 keV) or
hard band (2.0--10.0 keV) in Kohno et al. (2002) were missed in this study.
We also compared our results of
four intermediate- and high-mass YSOs (a$_s$, IRS1SW, IRS2, and IRS3NE)
with those in Kohno et al. (2002).
We do not report on X-rays from the position of a$_s$, although
Kohno et al. (2002) found a counterpart within a $1\arcsec$ radius.
We certainly found an X-ray source at a distance of $0\farcs54$ from a$_s$,
which is larger than our criterion radius of identification at the position of
a$_s$ of $0\farcs5$ based on the error radius derived
by the wavdetect program.
We, hence, do not regard this X-ray source as a counterpart of a$_s$.
For IRS1SW, IRS2, and IRS3NE, we obtained almost the same results
as that of Kohno et al. (2002).
Based on the positions in the color--color and color--magnitude diagrams
(subsections 4.3
and 4.4), IRS1SW is classified as a high- or intermediate-mass CTTS, and
IRS2 has a larger $K$-band excess than CTTSs. Because
IRS3NE is saturated at the $K$ band, the reliable $K$-band magnitude
can not be obtained.
However, based on the $J$- and $H$-band magnitudes, this is
classified as a high- or intermediate-mass source.

Preibisch et al. (2002) resolved a multiple IRS3 source into
three individual IR sources (IRS3A=IRS3S, IRS3C=IRS3NE, and IRS3E=CMDSH278)
by their deep observation using HST/NICMOS, and found that all of the sources
emit X-rays using the same dataset as ours.
We found five photons at the NIR position of IRS3S.
However, these photons may not be from IRS3
because of severe contamination from the bright source \#207
(=IRS3NE/IRS3C).
IRS3E, which corresponds to \#202, is classified as a low-mass CTTS, as also
shown in figure 2 in Preibisch et al. (2002).

\subsection{X-Ray Properties along the Evolutionary Phases}

\begin{figure}[htbp]
\begin{center}
\FigureFile(135mm,135mm){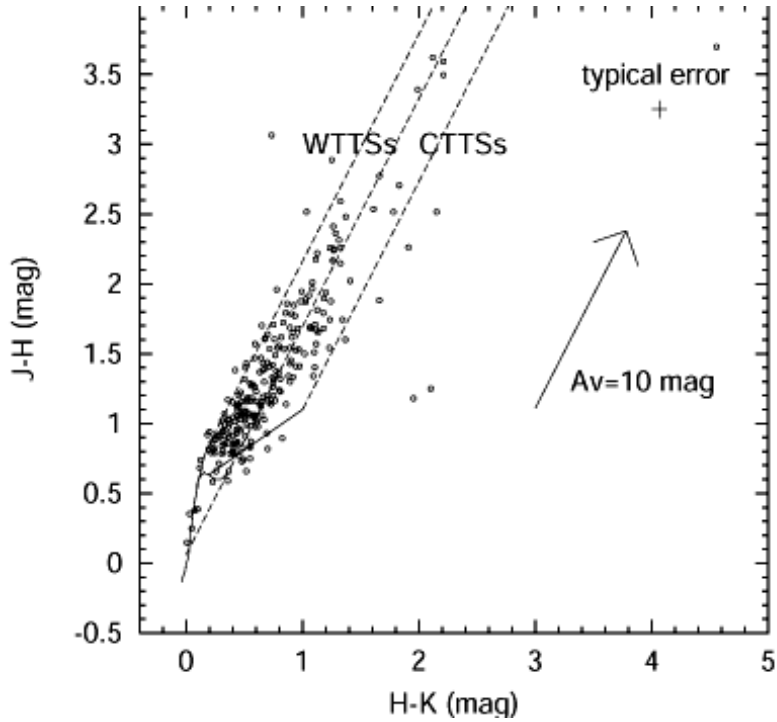}
\end{center}
\caption{$(J-H)/(H-K)$ color--color diagram in the CIT color system.
X-ray sources that have $J$-, $H$-, and $K$-band counterparts
are plotted with circles.
The thick solid line indicates the de-reddened CTTS locus (Meyer et~al. 1997).
The intrinsic colors of giants and dwarfs are given by the thin solid curves
(Tokunaga 2000).
Reddening lines are given in dash (Martin, Whittet 1990).}
\label{fig:colcol}
\end{figure}

We classified the 247 X-ray sources that have $J$-, $H$-,
and $K$-band counterparts (hereafter ``NIR-bright sources'')
into CTTS and WTTS by the following criterion.
Using the NIR data, we made the $(J-H)/(H-K)$ color--color diagram shown in 
figure \ref{fig:colcol}, where the colors were transformed into the CIT
(California Institute of Technology) system (Bessel, Brett 1988)
with a formula given by Carpenter (2001).
The thin solid curves are the intrinsic colors of giant and dwarf stars
(Tokunaga 2000), while the de-reddened locus (thick solid line)
for CTTSs was taken from Meyer et~al. (1997).
We assumed the slope of the reddening line to be $E(J-H)/E(H-K)$= 1.63
(Martin, Whittet 1990), as given by the dashed lines.
Since CTTSs are surrounded by cold
accretion disks, they generally have larger $K$-values than WTTSs.
Therefore, we regard the sources located between the middle and the right
reddening lines and above the CTTS locus as CTTSs,
while sources located between the left and middle reddening lines are
regarded as WTTSs.
We obtained 57 and 164 sources with this classification.
The remaining 26 sources located in other regions could not be classified.
The expected errors in the classification are presented by Meyer et al. (1997).
They concluded that approximately 1/3 of the CTTSs would occupy the region of
the WTTSs.
However, in this paper, we use the terms ``CTTSs" and ``WTTSs" for plainness.

Excluding the unclassified sources, the number of bright sources and
the averaged X-ray properties in each evolutionary phase are listed in table 3.
We can see a higher $kT$ for CTTSs (3.1 keV) than WTTSs (1.9 keV).
However, there should be a selection bias in favor of a less-absorbed WTTS.
Then, we also derived the averaged $kT$ for the sources that had a
$N_{\rm H}$ value less than 10$^{22}$ cm$^{-2}$.
Nevertheless, the $kT$ of CTTSs (2.4 keV) is still slightly higher than that
of WTTSs (2.0 keV).
The confidence level of the difference between the two distributions of $kT$ is
estimated to be 87.6\% using the ASURV statistical software package (ver.1.2).
\footnote{See $\langle$http://www.astro.psu.edu/statcodes/$\rangle$.}
In order to examine this feature for other SFRs,
we performed the same analysis using the published CXO catalogs of
the Orion Nebula Cluster (hereafter ONC, Feigelson et~al. 2002)
and the Orion Molecular Cloud-2/3 (hereafter OMC-2/3, Tsujimoto et~al. 2002),
other sites of high- and intermediate-mass star formation.
In identifying the 2MASS counterpart of X-ray sources in ONC, 
we defined the criterion that the angular separation
of X-ray and NIR pairs is smaller than the solid line
in figure \ref{fig:ang.sep.}, which is the envelope curve of the source
identification criterion for Mon R2.
For OMC-2/3, we used sources that
were already identified with 2MASS sources by Tsujimoto et~al. (2002).
We then classified the sources with the same criterion as applied for
Mon R2 and compared the mean X-ray temperature for each group, as shown
in table 3.
We again confirmed that the results similar to Mon R2; $kT$ of CTTSs is
slightly higher than that of WTTSs.
The confidence levels of difference between the two distributions of $kT$
are $>$~99.9\% and 46.2\%, respectively.
Although the significance level is not high enough, except for ONC,
we concluded that there is a marginal trend
that $kT$ of CTTSs is higher than that of WTTSs for all of the clouds
(Mon R2, ONC, and OMC-2/3).
This fact indicates that there is some changes between CTTSs and 
WTTSs in a physical process for the X-ray emission. 
In the case of YSO flares, Imanishi et~al. (2003) reported that
younger sources show a higher $kT$ than evolved sources
using a high-quality sample in the $\rho$ Ophiuchi low-mass SFR.
From a systematic study on the time profile of these flares,
they concluded that the magnetic field gradually
declines along the evolutionary phases, because the magnetic field strength has
a positive correlation with $kT$ (Shibata, Yokoyama 2002).
For Mon R2, we can not discuss the flare and quiescent phases separately
because of the limited photon counts.
Future observations with a deep exposure time will reveal whether
a similar effect can also be seen in Mon R2.

\subsection{X-ray Properties of Different Stellar Masses}

\begin{figure}[htbp]
\begin{center}
\FigureFile(100mm,100mm){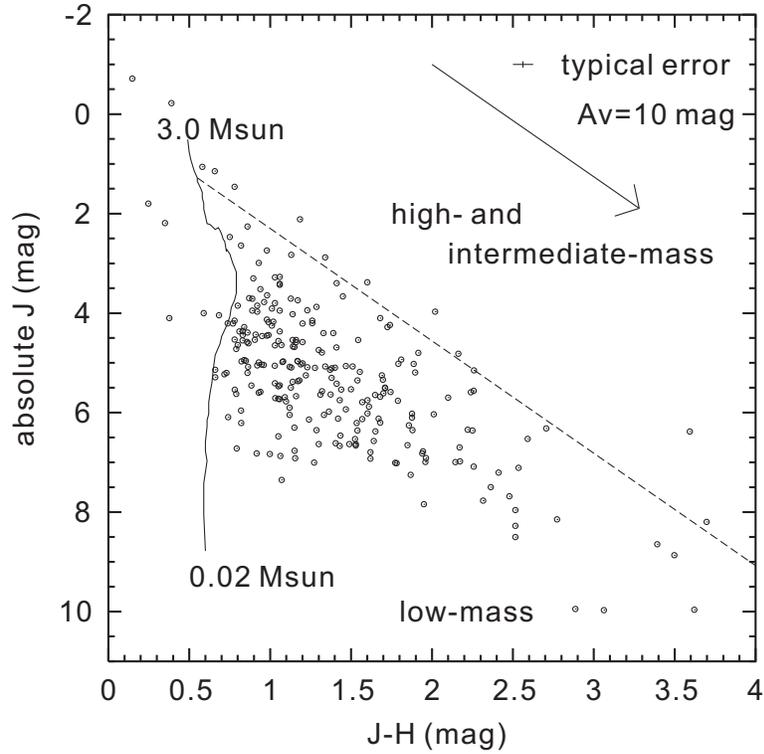}
\end{center}
\caption{Same as figure 6, but $J/(J-H)$ color--magnitude diagram.
The term ``absolute $J$" indicates the $J$-band magnitude of each source is
not corrected against the $A_V$ value.
The solid curve is the 1~Myr isochrone
with $0.02\ \MO < M < 3.0\ \MO$ (Baraffe et~al. 1998; Siess et~al. 2000).
The reddening line (dashed) is for stars with the
mass of $2.0\ \MO$ (Martin, Whittet 1990).}
\label{fig:magcol}
\end{figure}

We divided the NIR-bright sources into two mass ranges, defined as
$M >2.0\ \MO$ (high- and intermediate-mass), and $M < 2.0\ \MO$ (low-mass)
by the following criterion.
In the $J/(J-H)$ color--magnitude diagram (figure \ref{fig:magcol}),
the 1 Myr isochrone curve of $0.02\ \MO$ to $3.0\ \MO$ sources
(Baraffe et~al. 1998; Siess et~al. 2000) is shown by the solid line.
We assumed the slope of the reddening line to be $A_J/\it E(J-H)=$2.26
(Martin, Whittet 1990).
We regard the sources located above the reddening line of stars
with the mass of $2.0\ \MO$ (dashed line) and rightward from the isochrone
curve (solid curve) as high- and intermediate-mass sources,
while sources below the reddening line and rightward from the isochrone curve
to be low-mass sources.
We then classified 10 and 230 sources into high- and intermediate-mass sources,
and low-mass sources, respectively.
The seven sources located leftward of the isochrone curve
could not be classified.
Excluding the seven unclassified sources,
the number of bright sources and the averaged X-ray properties
of each mass range are given in table 4. The
$N\rm_H$ value for the low-mass sources are significantly smaller than that of
the high- and intermediate-mass sources.
This tendency may be due to the bias effect; the low-mass sources with large
absorption come more easily below the detection limit compared to those of
the high- and intermediate-mass sources, because the former has a
systematically lower $L\rm_X$ than the latter (Kohno et al. 2002).

Since we found a high $kT$ (2.7 keV) of the high- and intermediate-mass
sources similar to that of the low-mass sources (2.0 keV),
we also performed the same analysis for ONC and OMC-2/3,
and found similar results to Mon R2 (table 4).
Moreover, a significant fraction of the high- and intermediate-mass sources
show a rapid time variability.
These results indicate that the X-ray emitting process of high- and
intermediate-mass YSOs is the same as that of the low-mass sources,
i.e., magnetic activity (Kohno et al. 2002; Preibisch et al. 2002).
Because high- and intermediate-mass YSOs lack
surface convection layers (Palla, Stahler 1993),
the convective dynamo process may not work.
What kind of the magnetic amplification mechanism is working in high- and
intermediate-mass YSOs is still an open question.

\subsection{The Spatial Distribution of X-Ray Sources}

The spatial distribution of X-ray sources in Mon R2 seems to be
elongated in the same direction as the $^{13}$CO intensity structure,
from north to south (figure \ref{fig:acisI}).
We examined whether or not the X-ray and NIR source distributions
have the same structures as that of $^{13}$CO
by evaluating the correlation coefficient $R$ (Bevington, Robinson 1992).
We defined two source groups: those detected by X-rays and NIR.
We set the $x$- and $y$-axis along the edge of the ACIS-I FOV, and
calculated $R$ for each source group separately.
We then obtained $R$ = $-$0.19 and $-$0.04 for the X-ray detected group and
the NIR-detected group, respectively.
In order to estimate the significance level,
we randomly distributed the same number of test sources
as those of X-ray and NIR in the $x$--$y$ plane by 10000 times, and
evaluated $R$.
We then estimated the significance levels to be 3.7 $\sigma$ and 1.3 $\sigma$,
for the X-ray and NIR sources, respectively.
We also derived $R$ for sources that are detected by NIR, but not by X-ray,
and obtained $R$ = $-$0.03, the significance level of which is 1.0 $\sigma$.
Considering the significance levels, the distribution of the X-ray sources
has a more elongated structure than that of NIR.
This supports the observational review of Feigelson and Montmerle (1999)
asserting that X-rays are a homogeneous tracer of all stages of YSOs,
while the X-ray emission decreases
as sources are approaching to the main-sequence stage.
We confirmed this fact quantitatively for the first time by
comparing the spatial distribution of X-ray sources and NIR sources.

%%%%%%%%%%%%%%%%%%%%%%%%%%%%%%%%%%%%%
% Summary
%
\section{Summary}

We summarize the analysis of the 100~ks observation 
on Mon R2 with CXO ACIS-I as follows:
\begin{enumerate}

\item We detected 368 X-ray sources from the cloud and performed spectral
and timing analyses for most of the sources.
This provided one of the largest catalogs of X-ray sources in an SFR.

\item About 80\% of the detected X-ray sources have NIR counterparts in the
catalogs of 2MASS, SQIID, and IRCAM3.
From their $N\rm_H$ and $K$-band magnitudes,
we conclude that most of these NIR-identified
X-ray sources are likely cloud members.

\item We classified the cloud members into CTTSs and WTTSs
using a $(J-H)/(H-K)$ color--color diagram and compared
the fractions of the X-ray time variability, averaged absorption,
and temperature.
We found a marginal trend that
the temperatures of CTTSs are slightly higher than those of WTTSs.
We performed the same analysis for the ONC and OMC-2/3 members, and obtained
similar results as Mon R2.
Thus, the  physical structure of X-ray emission may change from CTTS to WTTS.

\item We divided the cloud members into high-, intermediate-,
and low-mass sources based on a $J/(J-H)$ color--magnitude diagram,
and compared the averaged X-ray properties.
A significant fraction of the high- and intermediate-mass sources
are time variable and exhibit a high-temperature plasma as those of
low-mass sources.
This supports the earlier results by Kohno et~al. (2002) and
Preibisch et al. (2002) that
high- and intermediate-mass YSOs emit X-rays
via magnetic activity.

\item We performed a imaging analysis of the entire cloud core
and the NIR cluster of Mon R2.
We then found a significant difference in the spatial distribution
between X-ray and NIR sources.
We confirmed that X-rays can be used as a homogeneous tracer of all the
stages of YSOs, as asserted by Feigelson and Montmerle (1999), and that
the X-ray emission decreases as sources are
approaching the main-sequence stage.

\end{enumerate}
%

%%%%%%%%%%%%%%%%%%%%%%%%%%%%%%%%%%%%%
% Reference
%

%%%%%%
% Figure 1
%
%%%%%%

%%%%%%
% Figure 2
%
%%%%%%

%%%%%%
% Figure 3
%
%%%%%%

%%%%%%
% Figure 4
%
%%%%%%

%%%%%%
% Figure 5
%
%%%%%%

%%%%%%
% Figure 6
%
%%%%%%%%%%%%

%%%%%%
% Figure 7
%
%%%%%%

%%%%%%
% Table 1
%
{\scriptsize
\begin{longtable}{cccr@{.}lcc}
\caption{Chandra X-ray sources and stellar counterparts.}
\hline \hline
ID & R.A.(J2000)\footnotemark[$*$] & Decl.(J2000)\footnotemark[$*$] & \multicolumn{2}{c}{Count\footnotemark[$\dagger$]} & \multicolumn{2}{c}{---------Identifications---------} \\
 & (mm ss) & (mm ss) &\multicolumn{2}{c}{}  & X-ray\footnotemark[$\ddagger$] & NIR\footnotemark[$\S$] \\
\hline
\endhead
\hline
\endfoot
\hline
\multicolumn{7}{l}{\hbox to 0pt{\parbox{125mm}{\footnotesize
\footnotemark[$*$] Right ascension and declination for all sources have the prefix of $\timeform{06h}$ and $\timeform{-06D}$ .
Coordinates are corrected to the 2MASS frame. \\
\footnotemark[$\dagger$] Background-subtracted X-ray photon counts in the 0.5--10.0 keV band. \\
\footnotemark[$\ddagger$] X-ray counterparts;
(KKH): Kohno et~al. (2002) and (GMCF): Gregorio-Hetem et~al. (1998). \\
\footnotemark[$\S$] NIR counterparts;
(CMDSH): Carpenter et al. (1997) and
(2MASSI): 2MASS Point Source Catalog in the Second Incremental Data Release.
NIR source names such as d$_{\rm N}$ and IRS4 are from Beckwith et al. (1976),
Howard et al. (1994), and Carpenter et al. (1997). \\
}}}
\endlastfoot
  1 & 07 13.48 & 20 06.7 &  13&6   & $\cdots$ & 2MASSI J0607134-062007 \\
  2 & 07 16.57 & 21 19.3 &  25&2   & $\cdots$ & $\cdots$ \\
  3 & 07 16.93 & 22 04.7 &  56&5   & $\cdots$ & 2MASSI J0607168-062205    \\
  4 & 07 17.69 & 19 43.4 &  54&8   & $\cdots$ & 2MASSI J0607177-061942    \\
  5 & 07 18.29 & 23 44.3 & 313&3   & GMCF8    & 2MASSI J0607182-062344    \\
  6 & 07 19.47 & 23 20.4 &  55&7   & $\cdots$ &   CMDSH1005 \\
  7 & 07 20.58 & 23 16.0 &  22&6   & $\cdots$ & $\cdots$ \\
  8 & 07 21.73 & 26 32.1 &  61&5   & $\cdots$ &   CMDSH1031 \\
  9 & 07 22.88 & 23 01.5 &  17&8   & $\cdots$ &   CMDSH1039 \\
 10 & 07 23.46 & 27 03.0 &  90&3   & $\cdots$ &   CMDSH1041 \\
 11 & 07 23.66 & 25 11.5 &  61&8   & $\cdots$ &   CMDSH1043 \\
 12 & 07 24.38 & 27 22.1 &  32&5   & $\cdots$ & $\cdots$ \\
 13 & 07 24.54 & 23 51.7 &  80&0   & $\cdots$ &   CMDSH1047 \\
 14 & 07 25.97 & 21 25.4 & 124&8   & $\cdots$ &   CMDSH1057 \\
 15 & 07 27.03 & 27 56.7 &  10&5   & $\cdots$ & $\cdots$ \\
 16 & 07 28.01 & 18 03.1 &  17&2   & $\cdots$ & $\cdots$ \\
 17 & 07 29.23 & 26 13.9 &  52&8   & $\cdots$ &   CMDSH1075 \\
 18 & 07 29.90 & 21 46.8 &  85&5   & $\cdots$ &   CMDSH1082 \\
 19 & 07 30.06 & 29 08.3 &  60&3   & $\cdots$ & $\cdots$ \\
 20 & 07 30.07 & 24 43.6 &  13&1   & $\cdots$ &   CMDSH1086
\end{longtable}
}
%%%%%%

%%%%%%
% Table 2
%
{\scriptsize
\begin{longtable}{ccr@{.}lr@{.}lcc}
\caption{Results of timing and spectral analyses for bright sources.}
\hline \hline
ID & Variable & \multicolumn{2}{c}{$N\rm_{H}$\footnotemark[$\dagger$]} & \multicolumn{2}{c}{$kT$\footnotemark[$\dagger$]\footnotemark[$\ddagger$]} & log($EM$\footnotemark[$\S$])\footnotemark[$\dagger$] & log($L\rm_{X}$\footnotemark[$\|$])\footnotemark[$\dagger$] \\
   & probability\footnotemark[$*$] & \multicolumn{2}{c}{($10^{22}$ cm$^{-2}$)} & \multicolumn{2}{c}{(keV)} & (cm$^{-3}$) & (erg s$^{-1}$) \\
\hline
\endhead
\hline
\endfoot
\hline
\multicolumn{8}{l}{\hbox to 0pt{\parbox{120mm}{\footnotesize
\footnotemark[$*$] Variable probabilities obtained by a Kolmogorov--Smirnov test to the photons in the 0.5--10.0 keV band. \\

\footnotemark[$\dagger$] 90\% confidence limits are in parentheses. No value in a parenthesis means that the error is not constrained. \\

\footnotemark[$\ddagger$] If sources have no converged value of $kT$, we assume that they have $kT$=10.0~keV. For \#347, the thin-thermal plasma model is rejected, while power-law model is accepted. \\

\footnotemark[$\S$] Emission measures. \\

\footnotemark[$\|$] Luminosity in the 0.5--10.0~keV band. \\

}}}
\endlastfoot
3   & 0.78   & 0&5$\;$(0.1--0.9)     & 0&7$\;$(0.5--1.1)      & 52.9$\;$(52.8--53.0) & 29.9$\;$(29.8--30.0) \\
4   & 0.99   & 1&0$\;$(...--...)   & 0&9$\;$(...--...)    & 53.1$\;$(53.0--53.2) & 30.1$\;$(30.0--30.2) \\
5   & 0.91   & 0&4$\;$(0.3--0.6)     & 1&1$\;$(1.0--1.3)      & 53.6$\;$(53.6--53.7) & 30.6$\;$(30.5--30.6) \\
6   & 0.73   & 1&0$\;$(...--...)   & 1&2$\;$(...--...)      & 53.1$\;$(53.0--53.2) & 30.1$\;$(30.0--30.2) \\
8   & 0.99   & 1&0$\;$(0.2--1.4)     & 0&6$\;$(0.4--1.7)      & 53.4$\;$(53.2--53.4) & 30.3$\;$(30.2--30.4) \\
10  & 0.99   & 1&5$\;$(0.6--2.0)     & 1&1$\;$(0.7--2.4)      & 53.5$\;$(53.4--53.6) & 30.5$\;$(30.4--30.5) \\
11  & 0.39   & 0&2$\;$(0.0--0.5)     & 4&0$\;$(...--37.8)     & 52.7$\;$(52.6--52.8) & 29.8$\;$(29.7--29.9) \\
13  & 0.98   & 1&9$\;$(1.0--3.5)     & 1&7$\;$(0.9--2.6)      & 53.4$\;$(53.3--53.5) & 30.4$\;$(30.3--30.5) \\
14  & 0.99   & 0&7$\;$(0.3--1.3)     & 4&7$\;$(1.8--...)     & 53.2$\;$(53.1--53.3) & 30.4$\;$(30.3--30.4) \\
17  & 0.99   & 1&3$\;$(...--...)   &26&9$\;$(...--...)    & 52.9$\;$(52.8--53.0) & 30.2$\;$(30.0--30.3) \\
18  & 0.96   & 0&4$\;$(0.2--0.8)     & 2&7$\;$(1.4--8.1)      & 53.0$\;$(52.9--53.1) & 30.0$\;$(29.9--30.1) \\
19  & 0.90   & 3&2$\;$(1.8--6.1)     & 7&9$\;$(...--...)     & 53.1$\;$(53.0--53.2) & 30.4$\;$(30.2--30.5) \\
21  & 0.70   & 0&2$\;$(0.1--0.5)     & 3&0$\;$(1.5--19.0)     & 52.8$\;$(52.7--52.9) & 29.9$\;$(29.8--30.0) \\
22  & 0.95   & 0&4$\;$(...--...)   & 2&3$\;$(...--...)     & 53.4$\;$(53.4--53.5) & 30.4$\;$(30.4--30.5) \\
25  & 0.70   & 0&9$\;$(...--...)   & 1&8$\;$(1.0--5.1)      & 53.0$\;$(52.9--53.1) & 30.0$\;$(29.9--30.1) \\
30  & 0.60   & 1&9$\;$(1.3--3.0)     & 1&9$\;$(1.2--3.7)      & 53.4$\;$(53.3--53.5) & 30.3$\;$(30.2--30.4) \\
32  & 0.89   & 2&9$\;$(2.2--...)    & 1&7$\;$(1.2--2.6)      & 53.9$\;$(53.8--54.0) & 30.8$\;$(30.8--30.9) \\
34  & 0.99   & 1&4$\;$(0.7--2.6)     & 5&1$\;$(1.9--...)      & 53.0$\;$(52.9--53.1) & 30.2$\;$(30.0--30.2) \\
37  & 0.92   & 0&6$\;$(0.3--1.6)     & 1&4$\;$(1.0--2.1)      & 53.0$\;$(52.9--53.1) & 29.9$\;$(29.8--30.0) \\
39  & 0.98   & 0&8$\;$(0.4--1.5)     & 1&4$\;$(1.1--2.4)      & 53.1$\;$(53.0--53.2) & 30.1$\;$(30.0--30.2) \\
42  & 1.00   & 0&5$\;$(0.4--...)    & 4&6$\;$(2.8--6.7)      & 53.3$\;$(53.3--53.4) & 30.5$\;$(30.4--30.5) \\
44  & 0.99   & 2&7$\;$(1.6--4.9)     & 5&6$\;$(1.6--...)      & 53.1$\;$(53.0--53.2) & 30.3$\;$(30.2--30.4) \\
54  & 0.99   & 3&7$\;$(2.6--5.3)     & 1&4$\;$(1.0--2.3)      & 53.8$\;$(53.7--53.9) & 30.7$\;$(30.6--30.8) \\
56  & 0.99   & 0&7$\;$(0.6--0.8)     & 2&2$\;$(1.8--2.7)      & 54.1$\;$(54.1--54.2) & 31.1$\;$(31.1--31.2) \\
58  & 0.71   & 2&5$\;$(1.7--3.6)     & 3&2$\;$(1.7--...)      & 53.6$\;$(53.5--53.6) & 30.7$\;$(30.6--30.7) \\
61  & 0.77   & 3&6$\;$(2.1--5.9)     &10&0                  & 53.2$\;$(53.1--53.3) & 30.3$\;$(30.2--30.4) \\
63  & 0.91   & 0&7$\;$(0.3--1.5)     & 5&1$\;$(2.0--55.7)     & 52.8$\;$(52.7--52.9) & 30.0$\;$(29.9--30.1) \\
64  & 0.52   & 0&4$\;$(0.1--0.7)     & 1&3$\;$(1.0--2.0)      & 52.8$\;$(52.6--52.9) & 29.7$\;$(29.6--29.8) \\
65  & 0.98   & 0&5$\;$(0.2--0.9)     & 2&7$\;$(1.3--10.3)     & 52.8$\;$(52.7--52.9) & 29.8$\;$(29.7--29.9) \\
66  & 0.81   & 4&9$\;$(2.8--15.7)    &13&7$\;$(...--...)     & 53.2$\;$(53.1--53.3) & 30.5$\;$(30.3--30.5) \\
67  & 0.99   & 0&9$\;$(0.1--1.4)     & 0&7$\;$(0.3--2.3)      & 53.2$\;$(53.1--53.3) & 30.2$\;$(30.1--30.3) \\
70  & 0.99   & 0&7$\;$(0.2--1.0)     & 1&1$\;$(0.9--1.8)      & 53.1$\;$(53.0--53.1) & 30.0$\;$(29.9--30.1) \\
71  & 0.75   & 0&6$\;$(0.2--1.1)     & 3&3$\;$(1.6--25.7)     & 52.9$\;$(52.8--52.9) & 29.9$\;$(29.8--30.0) \\
74  & 0.88   & 1&2$\;$(0.9--1.4)     & 0&6$\;$(0.4--0.8)      & 53.6$\;$(53.5--53.7) & 30.6$\;$(30.5--30.7) \\
78  & 0.99   & 2&7$\;$(1.6--4.3)     & 2&0$\;$(1.1--4.2)      & 53.4$\;$(53.3--53.5) & 30.4$\;$(30.3--30.5) \\
83  & 0.99   & 3&2$\;$(2.0--4.8)     & 1&7$\;$(1.0--3.1)      & 53.6$\;$(53.5--53.7) & 30.6$\;$(30.5--30.7) \\
105 & 0.89   & 0&2$\;$(0.1--0.4)     & 1&3$\;$(1.0--1.6)      & 53.3$\;$(53.3--53.4) & 30.3$\;$(30.2--30.3) \\
107 & 0.97   & 1&2$\;$(0.6--2.0)     & 2&4$\;$(1.1--7.5)      & 53.0$\;$(52.9--53.1) & 30.0$\;$(29.9--30.1) \\
122 & 0.99   & 6&8$\;$(5.3--9.1)     & 3&2$\;$(1.9--5.3)      & 54.1$\;$(54.1--54.2) & 31.2$\;$(31.2--31.3) \\
125 & 0.99   & 0&8$\;$(0.3--1.1)     & 0&9$\;$(0.6--1.2)      & 53.4$\;$(53.3--53.4) & 30.3$\;$(30.2--30.4) \\
126 & 0.96   & 2&3$\;$(1.5--3.6)     & 2&1$\;$(1.1--4.4)      & 53.4$\;$(53.3--53.5) & 30.4$\;$(30.3--30.5) \\
129 & 0.52   & 0&6$\;$(0.4--...)    & 1&8$\;$(1.3--2.0)      & 53.5$\;$(53.5--53.6) & 30.5$\;$(30.4--30.5) \\
131 & 0.99   & 0&7$\;$(0.4--1.0)     & 0&6$\;$(...--...)      & 53.2$\;$(53.0--53.2) & 30.1$\;$(30.0--30.2) \\
135 & 0.98   & 1&1$\;$(0.3--1.9)     & 1&0$\;$(0.4--3.0)      & 53.1$\;$(53.0--53.2) & 30.1$\;$(30.0--30.2) \\
141 & 0.87   & 2&0$\;$(1.8--2.4)     & 1&4$\;$(...--...)      & 54.2$\;$(54.2--54.3) & 31.2$\;$(31.1--31.2) \\
147 & 0.43   & 2&8$\;$(1.0--4.5)     & 1&1$\;$(0.7--1.7)      & 53.6$\;$(53.5--53.7) & 30.6$\;$(30.5--30.7) \\
149 & 0.89   & 0&8$\;$(0.3--1.5)     &10&0                  & 52.9$\;$(52.7--53.1) & 30.0$\;$(29.8--30.1) \\
150 & 0.95   & 0&2$\;$(0.1--0.4)     & 2&1$\;$(1.3--4.1)      & 53.0$\;$(53.0--53.1) & 30.0$\;$(30.0--30.1) \\
153 & 0.96   & 6&1$\;$(2.5--13.3)    & 2&9$\;$(...--...)      & 53.5$\;$(53.4--53.6) & 30.6$\;$(30.5--30.7) \\
154 & 0.53   & 0&2$\;$(0.1--0.7)     & 1&1$\;$(0.6--1.6)      & 52.7$\;$(52.6--52.8) & 29.6$\;$(29.5--29.7) \\
155 & 0.99   & 8&9$\;$(5.0--11.3)    & 6&9$\;$(...--...)      & 53.6$\;$(53.5--53.6) & 30.8$\;$(30.7--30.9) \\
156 & 0.98   & 1&7$\;$(1.5--2.0)     & 1&5$\;$(...--...)      & 54.2$\;$(54.2--54.2) & 31.2$\;$(31.1--31.2) \\
157 & 1.00   & 2&7$\;$(2.1--3.4)     & 5&1$\;$(3.3--13.6)     & 53.9$\;$(53.8--53.9) & 31.0$\;$(31.0--31.1) \\
160 & 0.97   &25&5$\;$(10.3--41.9)   & 1&8$\;$(0.9--5.8)      & 54.6$\;$(54.5--54.6) & 31.5$\;$(31.4--31.6) \\
161 & 0.77   & 0&2$\;$(0.0--1.0)     & 1&3$\;$(0.5--1.9)      & 52.6$\;$(52.5--52.7) & 29.6$\;$(29.5--29.7) \\
162 & 0.60   & 2&6$\;$(1.4--3.3)     & 2&9$\;$(1.8--10.4)     & 53.3$\;$(53.2--53.4) & 30.4$\;$(30.3--30.5) \\
165 & 0.99   & 8&9$\;$(5.6--15.6)    & 2&0$\;$(0.9--3.6)      & 54.2$\;$(54.1--54.2) & 31.2$\;$(31.1--31.2) \\
168 & 1.00   & 2&4$\;$(1.4--3.8)     & 5&3$\;$(2.2--26.0)     & 53.3$\;$(53.2--53.4) & 30.5$\;$(30.4--30.5) \\
169 & 0.98   & 3&9$\;$(2.5--6.0)     & 2&5$\;$(1.2--...)     & 53.7$\;$(53.6--53.8) & 30.7$\;$(30.7--30.8) \\
172 & 0.48   & 0&8$\;$(0.2--1.4)     & 1&4$\;$(0.9--3.3)      & 52.9$\;$(52.8--53.0) & 29.9$\;$(29.7--30.0) \\
179 & 0.99   & 6&5$\;$(3.6--11.3)    & 2&6$\;$(1.2--8.3)      & 53.7$\;$(53.6--53.8) & 30.8$\;$(30.7--30.8) \\
183 & 0.88   & 1&5$\;$(0.9--2.6)     & 2&6$\;$(1.5--6.9)      & 53.1$\;$(53.0--53.2) & 30.1$\;$(30.0--30.2) \\
184 & 0.94   & 0&7$\;$(0.2--1.7)     & 2&0$\;$(1.0--5.7)      & 52.8$\;$(52.7--52.9) & 29.8$\;$(29.7--29.9) \\
190 & 0.99   & 0&6$\;$(0.3--1.4)    & 13&0$\;$(...--...)      & 52.7$\;$(52.6--52.8) & 30.0$\;$(29.9--30.1) \\
192 & 0.99   & 0&4$\;$(0.2--0.8)     & 3&0$\;$(1.7--...)      & 53.2$\;$(53.2--53.3) & 30.3$\;$(30.2--30.4) \\
193 & 0.60   & 4&8$\;$(3.0--7.0)     & 2&1$\;$(1.1--4.3)      & 53.9$\;$(53.8--53.9) & 30.9$\;$(30.8--30.9) \\
197 & 1.00   &10&5$\;$(5.6--20.1)    & 2&0$\;$(...--8.3)      & 54.0$\;$(53.9--54.1) & 31.0$\;$(30.9--31.1) \\
198 & 1.00   & 1&7$\;$(1.2--2.5)     & 5&1$\;$(2.8--...)      & 53.6$\;$(53.5--53.6) & 30.7$\;$(30.7--30.8) \\
202 & 0.99   & 0&8$\;$(0.2--2.4)     & 3&9$\;$(...--...)      & 52.8$\;$(52.7--52.9) & 30.0$\;$(29.8--30.0) \\
209 & 1.00   & 2&7$\;$(2.5--2.9)     & 4&3$\;$(3.6--5.2)      & 54.7$\;$(54.7--54.7) & 31.8$\;$(31.8--31.9) \\
210 & 0.99   & 1&0$\;$(...--2.2)    & 3&4$\;$(...--62.0)     & 53.3$\;$(53.2--53.4) & 30.4$\;$(30.3--30.5) \\
212 & 0.97   & 5&0$\;$(3.2--...)    & 2&8$\;$(...--...)     & 53.9$\;$(53.9--54.0) & 31.0$\;$(30.9--31.0) \\
213 & 0.33   & 1&0$\;$(0.7--1.3)     & 2&0$\;$(...--...)     & 53.5$\;$(53.5--53.6) & 30.5$\;$(30.5--30.6) \\
217 & 0.99   & 0&3$\;$(0.2--0.4)     & 1&3$\;$(1.1--1.5)      & 53.4$\;$(53.3--53.4) & 30.3$\;$(30.3--30.4) \\
220 & 0.98   & 0&3$\;$(0.1--0.6)     & 1&9$\;$(1.3--2.8)      & 53.0$\;$(52.9--53.0) & 29.9$\;$(29.8--30.0) \\
224 & 1.00   & 2&4$\;$(1.4--3.6)     & 2&9$\;$(1.6--...)     & 53.6$\;$(53.5--53.7) & 30.7$\;$(30.6--30.7) \\
225 & 0.93   & 1&8$\;$(1.2--2.8)     & 3&6$\;$(1.9--10.7)     & 53.2$\;$(53.1--53.3) & 30.3$\;$(30.2--30.4) \\
226 & 0.95   & 1&2$\;$(0.4--1.8)     & 1&1$\;$(0.7--1.6)      & 53.3$\;$(53.2--53.4) & 30.2$\;$(30.1--30.3) \\
228 & 0.99   & 1&3$\;$(0.8--1.8)     & 1&1$\;$(0.8--1.5)      & 53.5$\;$(53.4--53.5) & 30.4$\;$(30.3--30.5) \\
229 & 0.97   & 0&7$\;$(0.4--1.3)     & 4&1$\;$(...--15.5)     & 53.1$\;$(53.0--53.2) & 30.3$\;$(30.2--30.4) \\
231 & 0.85   & 0&8$\;$(...--...)   & 0&9$\;$(...--...)     & 53.4$\;$(53.3--53.5) & 30.4$\;$(30.3--30.4) \\
237 & 0.91   & 1&8$\;$(0.7--2.6)     & 1&2$\;$(0.7--1.7)      & 53.4$\;$(53.3--53.5) & 30.3$\;$(30.2--30.4) \\
255 & 0.85   & 1&0$\;$(0.5--1.9)     & 2&2$\;$(1.1--4.6)      & 53.0$\;$(52.9--53.1) & 30.0$\;$(29.9--30.1) \\
256 & 0.79   & 2&2$\;$(1.5--3.5)     &10&0                  & 53.1$\;$(53.0--53.2) & 30.2$\;$(30.1--30.3) \\
259 & 0.92   & 0&8$\;$(0.2--...)    & 0&9$\;$(...--...)     & 53.5$\;$(53.4--53.6) & 30.5$\;$(30.4--30.5) \\
261 & 0.72   & 2&2$\;$(1.6--3.6)     & 1&7$\;$(0.9--3.2)      & 53.9$\;$(53.8--54.0) & 30.9$\;$(30.8--31.0) \\
264 & 0.98   & 1&2$\;$(0.5--2.6)     & 4&0$\;$(...--60.8)     & 53.0$\;$(52.9--53.1) & 30.1$\;$(30.0--30.2) \\
265 & 0.43   & 0&7$\;$(0.5--1.0)     & 0&7$\;$(...--...)     & 53.5$\;$(53.4--53.5) & 30.4$\;$(30.4--30.5) \\
268 & 0.71   & 1&2$\;$(0.5--1.6)     & 0&4$\;$(...--2.4)      & 53.7$\;$(53.6--53.8) & 30.6$\;$(30.5--30.7) \\
270 & 0.46   & 0&9$\;$(0.1--1.4)     & 1&0$\;$(0.5--11.0)     & 53.0$\;$(52.9--53.1) & 30.0$\;$(29.8--30.1) \\
277 & 0.99   & 2&9$\;$(...--...)   & 1&7$\;$(...--...)     & 53.7$\;$(53.7--53.8) & 30.7$\;$(30.6--30.8) \\
281 & 0.93   & 0&1$\;$(0.1--0.4)     & 4&8$\;$(1.4--...)     & 52.4$\;$(52.3--52.5) & 29.6$\;$(29.5--29.7) \\
286 & 0.97   & 1&2$\;$(0.8--2.1)     & 2&9$\;$(1.4--7.3)      & 53.2$\;$(53.1--53.3) & 30.2$\;$(30.1--30.3) \\
289 & 0.94   & 1&2$\;$(...--1.4)    & 1&1$\;$(...--...)     & 54.2$\;$(54.2--54.2) & 31.1$\;$(31.1--31.2) \\
293 & 0.95   & 0&4$\;$(0.2--0.9)     & 4&2$\;$(2.0--12.3)     & 52.8$\;$(52.7--52.9) & 30.0$\;$(29.9--30.1) \\
295 & 0.84   & 2&2$\;$(1.5--3.3)     &10&0                  & 53.1$\;$(52.9--53.2) & 30.2$\;$(30.0--30.3) \\
296 & 1.00   & 0&3$\;$(0.1--0.4)     & 2&3$\;$(1.6--3.3)      & 53.2$\;$(53.2--53.3) & 30.2$\;$(30.2--30.3) \\
301 & 0.97   & 0&5$\;$(0.1--...)    & 0&7$\;$(0.5--1.1)      & 53.5$\;$(53.5--53.6) & 30.5$\;$(30.4--30.5) \\
304 & 0.99   & 0&6$\;$(0.4--...)    & 2&9$\;$(2.3--3.4)      & 53.9$\;$(53.9--53.9) & 31.0$\;$(30.9--31.0) \\
305 & 0.94   & 1&4$\;$(1.0--1.9)    & 20&2$\;$(...--...)     & 53.8$\;$(53.8--53.8) & 31.1$\;$(31.0--31.1) \\
307 & 0.82   & 0&6$\;$(0.1--1.1)     & 0&9$\;$(0.6--1.7)      & 53.0$\;$(52.9--53.1) & 30.0$\;$(29.9--30.1) \\
309 & 0.99   & 0&3$\;$(0.1--0.7)     & 3&4$\;$(1.7--8.3)      & 52.8$\;$(52.7--52.8) & 29.9$\;$(29.8--29.9) \\
310 & 0.99   & 0&5$\;$(...--...)   & 1&6$\;$(1.3--1.8)      & 54.2$\;$(54.1--54.2) & 31.1$\;$(31.1--31.1) \\
311 & 0.99   & 0&6$\;$(...--...)   & 1&5$\;$(1.3--1.8)      & 54.0$\;$(54.0--54.1) & 31.0$\;$(31.0--31.0) \\
313 & 0.94   & 0&2$\;$(0.1--0.4)     & 1&6$\;$(1.0--2.9)      & 53.2$\;$(53.1--53.3) & 30.2$\;$(30.1--30.3) \\
317 & 0.99   & 0&6$\;$(0.4--...)    & 1&9$\;$(1.1--2.4)      & 53.4$\;$(53.4--53.5) & 30.4$\;$(30.3--30.4) \\
320 & 0.99   & 0&2$\;$(0.2--0.4)     & 2&1$\;$(1.6--...)     & 53.5$\;$(53.5--53.6) & 30.5$\;$(30.5--30.6) \\
321 & 0.92   & 1&4$\;$(0.9--2.1)     & 4&9$\;$(...--11.5)     & 53.4$\;$(53.4--53.5) & 30.6$\;$(30.5--30.7) \\
329 & 0.78   & 0&4$\;$(0.2--0.7)     & 1&3$\;$(1.0--1.7)      & 53.1$\;$(53.0--53.1) & 30.0$\;$(29.9--30.1) \\
331 & 0.82   & 3&6$\;$(2.6--5.5)    & 23&1$\;$(...--...)      & 53.7$\;$(53.6--53.7) & 31.0$\;$(30.9--31.0) \\
333 & 0.99   & 0&3$\;$(0.1--0.8)     & 1&5$\;$(0.8--2.5)      & 53.2$\;$(53.1--53.3) & 30.2$\;$(30.1--30.2) \\
344 & 0.96   & 0&2$\;$(0.1--0.9)     & 1&2$\;$(0.5--1.6)      & 52.8$\;$(52.7--52.9) & 29.8$\;$(29.7--29.8) \\
347 & 1.00   & 0&2$\;$(0.1--0.3)     & \multicolumn{2}{c}{...}&        ...           & 29.8$\;$(29.7--29.8) \\
349 & 0.99   & 1&7$\;$(1.1--2.8)     & 0&9$\;$(0.5--1.9)      & 53.4$\;$(53.3--53.5) & 30.4$\;$(30.2--30.5) \\
350 & 1.00   & 1&3$\;$(1.2--...)    & 4&3$\;$(3.6--5.2)      & 54.5$\;$(54.5--54.6) & 31.7$\;$(31.7--31.7) \\
351 & 0.99   & 0&6$\;$(0.2--1.1)     & 0&9$\;$(0.7--1.4)      & 53.1$\;$(53.0--53.2) & 30.0$\;$(29.9--30.1) \\
353 & 0.60   & 1&0$\;$(0.7--1.5)     &10&0                  & 53.2$\;$(53.1--53.3) & 30.3$\;$(30.2--30.4) \\
354 & 0.99   & 1&0$\;$(0.2--1.6)     & 1&0$\;$(0.7--1.7)      & 53.2$\;$(53.1--53.3) & 30.1$\;$(30.0--30.2) \\
355 & 0.56   & 0&8$\;$(0.5--1.1)     & 0&5$\;$(...--...)      & 53.3$\;$(53.1--53.4) & 30.2$\;$(30.1--30.3) \\
357 & 0.88   & 4&2$\;$(2.6--6.6)     &10&0                  & 53.2$\;$(53.0--53.4) & 30.3$\;$(30.1--30.5) \\
358 & 0.99   & 0&8$\;$(...--...)   & 1&0$\;$(0.8--2.3)      & 53.7$\;$(53.7--53.8) & 30.7$\;$(30.6--30.7) \\
359 & 0.96   & 1&0$\;$(0.5--2.0)    & 28&3$\;$(...--...)      & 52.9$\;$(52.8--53.0) & 30.2$\;$(30.0--30.2) \\
360 & 1.00   & 0&5$\;$(0.3--0.9)     & 3&7$\;$(...--22.7)     & 52.8$\;$(52.7--52.9) & 30.0$\;$(29.9--30.1) \\
363 & 0.99   & 2&2$\;$(1.9--2.5)     & 2&8$\;$(...--3.5)      & 54.4$\;$(54.3--54.4) & 31.4$\;$(31.4--31.4) \\
366 & 0.63   & 0&8$\;$(0.7--...)    & 1&7$\;$(1.3--2.0)      & 54.1$\;$(54.1--54.1) & 31.0$\;$(31.0--31.1) \\
367 & 0.91   & 0&6$\;$(0.3--1.4)     & 2&1$\;$(1.0--3.1)      & 53.2$\;$(53.1--53.3) & 30.2$\;$(30.1--30.3) \\
\end{longtable}
}
%%%%%%

%%%%%%
% Table 3
%
\begin{table}
\begin{center}
\caption{X-ray properties along the evolutionary phases.\label{tab:class}}
\begin{tabular}{p{4cm}ll}
\hline \hline
     & CTTSs & WTTSs \\ \hline
     Mon R2 & & \\
    Number\footnotemark[$*$]  &  23 (12) & 69 (45) \\
    Variability\footnotemark[$\dagger$] (ratio $\%$) &  8  (34.8) & 23  (33.3) \\
    $N_{\rm H}$ ($10^{22}$ cm$^{-2}$)\footnotemark[$\ddagger$] &  1.3  (1.0)  & 0.9  (0.8)  \\
    $kT_1$ (keV)\footnotemark[$\ddagger$] & 3.1  (2.5)  & 1.9  (1.4)  \\
    $kT_2$ (keV)\footnotemark[$\ddagger$] & 2.4  (1.1)  & 2.0  (1.5)  \\
\hline
     ONC & & \\
    Number\footnotemark[$*$] &  65  (56) & 285  (250) \\
    $kT_1$ (keV)\footnotemark[$\ddagger$] & 2.9  (1.5)  & 2.3  (1.4)  \\
    $kT_2$ (keV)\footnotemark[$\ddagger$] & 3.1  (1.6)  & 2.4  (1.4)  \\
\hline
     OMC-2/3 & &\\
    Number\footnotemark[$*$] &  14  (7) & 92  (70) \\
    $kT_1$ (keV)\footnotemark[$\ddagger$] & 2.1  (1.1) & 1.4  (0.7)  \\
    $kT_2$ (keV)\footnotemark[$\ddagger$] & 1.5  (0.7) & 1.3  (0.7)  \\
\hline
\multicolumn{3}{@{}l@{}}{\hbox to 0pt{\parbox{85mm}{\footnotesize
    \footnotemark[$*$] The number of sources that have $N_{\rm H}$ value of less than $10^{22}$ cm$^{-2}$ are given in parentheses.
In the case of Mon R2, the number of the bright sources in each evolutionary phase are shown. \\
    \footnotemark[$\dagger$] The number of the time-variable sources. The
fraction of the number of time-variable sources are given in parentheses. \\
    \footnotemark[$\ddagger$] The average of the best fit values (not weighted
by the errors). The standard deviations are given in parentheses. \\
 }\hss}}
\end{tabular}
\end{center}
\end{table}
%%%%%%

%%%%%%
% Table 4
%
\begin{table}
\begin{center}
\caption{X-ray properties of different stellar masses.\label{tab:mass}}
\begin{tabular}{p{4cm}p{1.9cm}l}
\hline \hline
        &High- and                   & Low mass   \\
        &intermediate-mass           &            \\ \hline
     Mon R2 & & \\
    Number\footnotemark[$*$]  &  8 & 92 \\
    Variability\footnotemark[$\dagger$] (ratio $\%$) &  3  (37.5) & 33  (35.9) \\
    $N_{\rm H}$ ($10^{22}$ cm$^{-2}$)\footnotemark[$\ddagger$] & 2.6  (2.6) & 1.0  (0.8) \\
    $kT$ (keV)\footnotemark[$\ddagger$] &  2.7  (1.9) & 2.0  (1.4)  \\
\hline
     ONC & & \\
    Number &  14 & 372 \\
    $kT$ (keV)\footnotemark[$\ddagger$] &  3.1  (1.4) & 2.4  (1.4) \\
\hline
     OMC-2/3 & & \\
    Number &  9 & 94 \\
    $kT$ (keV)\footnotemark[$\ddagger$] &  2.1  (1.2) & 1.4  (1.2) \\
\hline
\multicolumn{3}{@{}l@{}}{\hbox to 0pt{\parbox{85mm}{\footnotesize
    \footnotemark[$*$] The number of the bright sources in each mass range. \\
    \footnotemark[$\dagger$] The number of the time-variable sources.
The fraction of the number of time-variable sources are given in parentheses. \\
    \footnotemark[$\ddagger$] The average of the best-fit values (not
weighted by the errors). The standard deviations are given in parentheses. \\
 }\hss}}
\end{tabular}
\end{center}
\end{table}
%%%%%%

\end{document}